\documentclass[showpacs,amssymb,amsmath,floatfix,prd,twocolumn]{revtex4}

\def\Ri2{R_{\mu\nu}R^{\mu\nu}}

\begin{document}

\title{Anisotropically Inflating Universes}
\author{John D. Barrow$^*$  and Sigbj\o rn Hervik$^{\dagger}$}
\affiliation{{}$^*$DAMTP, Centre for Mathematical Sciences, Cambridge University, 
Wilbeforce Road, Cambridge CB3 0WA, UK. \\
{}$^\dagger$Department of Mathematics and Statistics, Dalhousie University,
Halifax, Nova Scotia, B3H 3J5 Canada. \\
\upshape E-mail: \texttt{J.D.Barrow@damtp.cam.ac.uk}, \texttt{herviks@mathstat.dal.ca}}
\date{\today }

\begin{abstract}
We show that in theories of gravity that add quadratic curvature invariants
to the Einstein-Hilbert action there exist expanding vacuum cosmologies with
positive cosmological constant which do not approach the de Sitter universe.
Exact solutions are found which inflate anisotropically. This behaviour is
driven by the Ricci curvature invariant and has no counterpart in the
general relativistic limit. These examples show that the cosmic no-hair
theorem does not hold in these higher-order extensions of general relativity
and raises new questions about the ubiquity of inflation in the very early
universe and the thermodynamics of gravitational fields.
\end{abstract}

\pacs{95.30.Sf, 98.80.Jk, 04.80.Cc, 98.80.Bp, 98.80.Ft, 95.10.Eg}
\maketitle



\section{Introduction}

The inflationary universe is the central cosmological paradigm which
astronomical observations aim to test, and by which we seek to understand
how the universe might have evolved from a general initial condition into
its present state of large-scale isotropy and homogeneity together with an
almost flat spectrum of near-Gaussian fluctuations. The essential feature of
this inflationary picture is a period of accelerated expansion during the
early stages of the universe \cite{inflate}. The simplest
physically-motivated inflationary scenario drives the acceleration by a
scalar field with a constant potential, and the latter can also be described
by adding a positive cosmological constant to the Einstein equations. In
order to understand the generality of this scenario it is important to
determine whether universal acceleration and asymptotic approach to the de
Sitter metric always occurs. A series of cosmic no-hair theorems of varying
strengths and degrees of applicability have been proved to demonstrate some
necessary and sufficient conditions for its occurrence \cite{jdb, bou,
starob, jss, wald}. Similar deductions are possible for power-law \cite{jb4,
power} and intermediate inflationary behaviour, \cite{inter}, where
accelerated expansion is driven by scalar-field potentials that have slow
exponential or power-law fall-offs, but we will confine our discussion to
the situation that occurs when there is a positive cosmological constant, $%
\Lambda >0$. So far, investigations have not revealed any strong reason to
doubt that, when $\Lambda >0$ and other matter is gravitationally
attractive, any stable, ever-expanding general-relativistic cosmological
model will approach isotropic de Sitter inflation exponentially rapidly
within the event horizon of any geodesically-moving observer. Similar
conclusions result when we consider inflation in those generalisations of
general relativity in which the Lagrangian is a function only of the scalar
curvature, $R$, of spacetime. This similarity is a consequence of the
conformal equivalence between these higher-order theories in vacuum and
general relativity in the presence of a scalar field \cite{barr, barrcot,
maeda}. In this paper we will show that when quadratic terms formed from the
Ricci curvature scalar, $R_{\mu \nu }R^{\mu \nu }$ are added to the
Lagrangian of general relativity then new types of cosmological solution
arise when $\Lambda >0$ which have no counterparts in general relativity.
They inflate anisotropically and do not approach the de Sitter spacetime at
large times. We give two new exact solutions for spatially homogeneous
anisotropic universes with $\Lambda >0$ which possess this new behaviour.
They provide counter-examples to the expectation that a cosmic no-hair
theorem will continue to hold in simple higher-order extensions of general
relativity. Other consequences of such higher-order theories have been studied 
in \cite{Berkin,BCX,Schmidt}. The presence of such quadratic terms as classical or quantum
corrections to the description of the gravitational field of the very early
universe will therefore produce very different outcomes following expansion
from general initial conditions to those usually assumed to arise from
inflation. This adds new considerations to the application of the chaotic
and eternal inflationary theories \cite{eternal} in conjunction with
anthropic selection \cite{BT}.

We will consider a theory of gravity derived from an action quadratic in the
scalar curvature and the Ricci tensor. More specifically, ignoring the
boundary term, we will consider the $D$-dimensional gravitational action 
\begin{equation}
S_{G}=\frac{1}{2\kappa }\int_{M}\mathrm{d}^{D}x\sqrt{|g|}\left( R+\alpha
R^{2}+\beta \Ri2-2\Lambda \right) .  \label{a}
\end{equation}%
Variation of this action leads to the following generalised Einstein
equations (see, e.g., \cite{DT}): 
\begin{equation}
G_{\mu \nu }+\Phi _{\mu \nu }+\Lambda g_{\mu \nu }=\kappa T_{\mu \nu },
\label{field}
\end{equation}%
where $T_{\mu \nu }$ is the energy-momentum tensor of the ordinary matter
sources, which we in this paper will assume to be zero, and 
\begin{eqnarray}  \label{c}
\lefteqn{G_{\mu \nu } \equiv R_{\mu \nu }-\frac{1}{2}Rg_{\mu \nu },}
\label{b} \\
\lefteqn{\Phi _{\mu \nu } \equiv}  \nonumber \\
&& 2\alpha R\left( R_{\mu \nu }-\frac{1}{4}Rg_{\mu \nu }\right) +(2\alpha
+\beta )\left( g_{\mu \nu }\Box -\nabla _{\mu }\nabla _{\nu }\right) R 
\nonumber \\
&&+\beta \Box \left( R_{\mu \nu }-\frac{1}{2}Rg_{\mu \nu }\right) +2\beta
\left( R_{\mu \sigma \nu \rho }-\frac{1}{4}g_{\mu \nu }R_{\sigma \rho
}\right) R^{\sigma \rho },  \nonumber \\
\end{eqnarray}%
with $\Box \equiv \nabla ^{\mu }\nabla _{\mu }$. The tensor $\Phi _{\mu \nu
} $ incorporates the deviation from regular Einstein gravity, and we see
that $\alpha =\beta =0$ implies $\Phi _{\mu \nu }=0$.

First, consider an Einstein metric, so that $R_{\mu \nu }=\lambda g_{\mu \nu
}$. This is a solution of eq.(\ref{field}) with $T_{\mu \nu }=0$ provided
that 
\begin{equation}
\Lambda =\frac{\lambda }{2}\left[ (D-4)(D\alpha +\beta )\lambda+(D-2)\right]
.  \label{d}
\end{equation}%
Hence, when $D=4$ any Einstein space is a solution to eq.(\ref{field})
provided that $\Lambda =(D-2)\lambda /2$. In particular, if $\Lambda >0$, de
Sitter spacetime is a solution to eq.(\ref{field}). If $\Lambda =0$, we need $%
\lambda =0$ and de Sitter spacetime cannot be a solution.

Now consider solutions to eq.(\ref{field}) which are non-perturbative and $%
\alpha $ and $\beta $ are not small. We know that solutions with $\beta
=0,\alpha \neq 0$ are conformally related to Einstein gravity with a scalar
field $\phi =\ln (1+2\alpha R)$ that possesses a self-interaction potential
of the form\emph{\ }$V(\phi )=(e^{\phi }-1)^{2}/4\alpha $, \cite{barr,
barrcot, maeda}, and their inflationary behaviours for small and large $%
\left\vert \phi \right\vert $, along with that of theories derived \ from
actions that are arbitrary functions of $R$, are well understood. However,
there is no such conformal equivalence with general relativity when $\beta
\neq 0$ and cosmologies with $\Lambda >0$ can then exhibit quite different
behaviour.

\section{The flat de Sitter solution}

First consider the spatially-flat de Sitter universe with metric 
\begin{equation}
\mathrm{d}s_{\text{dS}}^{2}=-\mathrm{d}t^{2}+e^{2Ht}\left( \mathrm{d}x^{2}+%
\mathrm{d}y^{2}+\mathrm{d}z^{2}\right) ,\quad H=\sqrt{\frac{\Lambda }{3}}.
\label{deS}
\end{equation}%
The stability of this solution in terms of perturbations of the scale-factor
depends on the sign of $(3\alpha +\beta )$. In 4D, we can use the Weyl
invariant and the Euler density, $E$, defined by \footnote{%
These expressions are only valid in 4D.}, 
\begin{eqnarray}
C_{\mu \nu \rho \sigma }C^{\mu \nu \rho \sigma } &=&R_{\mu \nu \rho \sigma
}R^{\mu \nu \rho \sigma }-2R_{\mu \nu }R^{\mu \nu }+\frac{1}{3}R^{2}, 
\nonumber \\
E &=&R_{\mu \nu \rho \sigma }R^{\mu \nu \rho \sigma }-4R_{\mu \nu }R^{\mu
\nu }+R^{2},
\end{eqnarray}%
to eliminate the quadratic Ricci invariant in the action, since 
\[
\alpha R^{2}+\beta R_{\mu \nu }R^{\mu \nu }=\tfrac{1}{3}(3\alpha +\beta
)R^{2}+\tfrac{\beta }{2}\left( C_{\mu \nu \rho \sigma }C^{\mu \nu \rho
\sigma }-E\right) . 
\]%
Since integration over the Euler density is a topological invariant, the
variation of $E$ will not contribute to the equations of motion. The
Friedmann-Robertson-Walker (FRW) universes are conformally flat so, for a
small variation, the invariant $C_{\mu \nu \rho \sigma }C^{\mu \nu \rho
\sigma }$ will not contribute either. Hence, sufficiently close to a FRW
metric only the $R^{2}$ term will contribute. The stability of the FRW
universe is therefore determined by the sign of $(3\alpha +\beta )$ \cite{BO}%
. One can check this explicitly using eq. (\ref{field}). We start with the
metric ansatz: 
\[
\mathrm{d}s^{2}=-\mathrm{d}t^{2}+e^{2b(t)}\left( \mathrm{d}x^{2}+\mathrm{d}%
y^{2}+\mathrm{d}z^{2}\right) ,\quad H=\sqrt{\frac{\Lambda }{3}}, 
\]%
and note that in 4D the trace of eq.(\ref{field}) reduces to 
\begin{equation}
-R+2(3\alpha +\beta )\Box R+4\Lambda =0,  \label{tr}
\end{equation}%
which can be used to determine the stability of the Ricci scalar. We can
perturb the Ricci scalar by assuming a small deviation from the flat de
Sitter metric of the form: 
\[
b(t)=Ht+b_{1}e^{\lambda _{1}t}+b_{2}e^{\lambda _{2}t}+\mathcal{O}%
(e^{2\lambda _{i}t}), 
\]%
where $b_{1}$ and $b_{2}$ are arbitrary constants. Eq.(\ref{tr}) implies 
\begin{equation}
\lambda _{1,2}=-\frac{3H}{2}\left( 1\pm \sqrt{1-\frac{2}{9H^{2}(3\alpha
+\beta )}}\right) ,  \label{eig}
\end{equation}%
if $(3\alpha +\beta )\neq 0$. For $(3\alpha +\beta )=0$, we must have $%
b_{1}=b_{2}=0$. From this expression we see that if $(3\alpha +\beta )>0$
then the solution will asymptotically approach the flat de Sitter spacetime
as $t\rightarrow \infty $; however, for $(3\alpha +\beta )<0$ the solution
is unstable. For the special case of $\beta =0$, this result agrees with the
stability analysis of \cite{BO}. A construction of an asymptotic series
approximation around the de Sitter metric for the case $\beta =0$\ has also been
performed \cite{st, bm, gott, ss, gss}. In the case of general relativity ($%
\alpha =\beta =0$) a number of results for the inhomogeneous case of small
perturbations from isotropy and homogeneity when $\Lambda >0$ have also been
obtained \cite{jdb, bou, starob, jss, bg, rend, Faraoni,FN}.\emph{\ }

We see\ that, as long as $(3\alpha +\beta )>0,$ any FRW model sufficiently
close to the flat de Sitter model will asymptotically approach de Sitter
spacetime and consequently obeys the cosmological no-hair theorem. We should
emphasize that only FRW perturbation modes have been considered here. The
question of whether the flat de Sitter universe is stable against general
anisotropic or large inhomogeneous perturbations when $\alpha \neq 0$ and $%
\beta \neq 0$ is still unsettled. In the case of universes that are not
'close' to isotropic and homogeneous FRW models we shall now show that the
cosmic no-hair theorem for $\Lambda >0$ vacuum cosmologies is not true:
there exist ever-expanding vacuum universes with $\Lambda >0$ that do not
approach the de Sitter spacetime.

\section{Exact anisotropic solutions}

We now present two new classes of exact vacuum anisotropic and spatially
homogeneous universes of Bianchi types $II$ and $VI_{h}$ with $\Lambda >0$.
These are new exact solutions of the eqns. (\ref{field}) with $(\alpha
,\beta )\neq (0,0)$.

\textbf{Bianchi type $II$ solutions:}

\begin{equation}
\mathrm{d}s_{{II}}^{2}=-\mathrm{d}t^{2}+e^{2bt}\left[ \mathrm{d}x+\frac{a}{2}%
(z\mathrm{d}y-y\mathrm{d}z)\right] ^{2}+e^{bt}(\mathrm{d}y^{2}+\mathrm{d}%
z^{2}),  \label{type2}
\end{equation}%
where 
\begin{equation}
a^{2}=\frac{11+8\Lambda (11\alpha+3\beta )}{30\beta },\quad b^{2}=\frac{%
8\Lambda (\alpha+3\beta )+1}{30\beta }.  \label{type2a}
\end{equation}

These solutions are spacetime homogeneous with a 5-dimensional isotropy
group. They have a one-parameter family of 4-dimensional Lie groups, as well
as an isolated one (with Lie algebras $A_{4,11}^{q}$ and $A_{4,9}^{1}$,
respectively, in Patera et al's scheme \cite{PSWZ}) acting transitively on
the spacetime. An interesting feature of this family of solutions is that
there is a lower bound on the cosmological constant, given by $\Lambda _{%
\text{min}}=-1/[8(\alpha +3\beta )]=-a^{2}/8$ for which the spacetime is
static. For $\Lambda >\Lambda _{\text{min}}$ the spacetime is inflating and
shearing. The inflation does not result in approach to isotropy or to
asymptotic evolution close to the de Sitter metric. Interestingly, even in
the case of a vanishing $\Lambda $ the universe inflates exponentially but
anisotropically. We also note from the solutions that the essential term in
the action causing this solution to exist is the $\beta R_{\mu \nu }R^{\mu
\nu }$-term and the distinctive behaviour occurs when $\alpha =0$. The
solutions have no well defined $\beta \rightarrow 0$ limit, and do not have
a general relativistic counterpart. They are non-perturbative. Similar
solutions exist also in higher dimensions. Their existence seem to be
related to so-called Ricci nilsolitons \cite{L1,SigTalk}.

\textbf{Bianchi type $VI_{h}$ solutions:}

\begin{eqnarray}
\lefteqn{\mathrm{d}s_{{VI}_{h}}^{2}=-\mathrm{d}t^{2}+\mathrm{d}x^{2}} 
\nonumber \\
&& +e^{2(r t+ax)}\left[ e^{-2(s t+a\tilde{h}x)}\mathrm{d}y^{2}+e^{+2(s t+a%
\tilde{h}x)}\mathrm{d}z^{2}\right] ,  \label{type6}
\end{eqnarray}%
where 
\begin{eqnarray}
r^{2}&=&\frac{8\beta s^2+(3+\tilde{h}^{2})(1+8\Lambda \alpha )+8\Lambda
\beta (1+\tilde{h}^{2})}{8\beta \tilde{h}^{2}},  \nonumber \\
a^{2}&=&\frac{8\beta s ^{2}+8\Lambda (3\alpha+\beta)+3}{8\beta \tilde{h}^{2}}%
.  \label{type6a}
\end{eqnarray}%
and $r $, $s$, $a$, and $\tilde{h}$ are all constants. These are also
homogeneous universes with a 4-dimensional group acting transitively on the
spacetime. Both the mean Hubble expansion rate and the shear are constant.
Again, we see that the solution inflates anisotropically and is supported by
the existence of $\beta \neq 0$. It exists when $\alpha =0$ and $\Lambda =0$
but not in the limit $\beta \rightarrow 0$. \emph{\ }

\section{Avoidance of the no-hair theorem}

The no-hair theorem for Einstein gravity states that for Bianchi types $%
I-VIII$ the presence of a positive cosmological constant drives the
late-time evolution towards the de Sitter spacetime. An exact statement of
the theorem can be found in the original paper by Wald \cite{wald}. It
requires the matter sources (other than $\Lambda $) to obey the
strong-energy condition. It has been shown that if this condition is relaxed
then the cosmic no-hair theorem cannot be proved and counter-examples exist 
\cite{jb1, jb2, jb3, jb4}. 
In \cite{Kaloper}, the cosmic no-hair conjecture was discussed for
Bianchi cosmologies with an axion field with a Lorentz Chern-Simons
term. Interestingly, exact Bianchi type $II$ solutions, similar to the
ones found here, were found which avoided  the cosmic no-hair
theorem. However, unlike for our solutions, these violations were
driven by an axion  field whose energy-momentum tensor violated the
strong and dominant energy condition.   
The no-hair theorem for spatially homogeneous
solutions of Einstein gravity also requires the spatial 3-curvature to be
non-positive. This condition ensures that universes do not recollapse before
the $\Lambda $ term dominates the dynamics but it also excludes examples
like that of the Kantowski-Sachs $S^{2}\times S^{1}$ universe which has an
exact solution with $\Lambda >0$ which inflates in some directions but is
static in others. These solutions, found by Weber \cite{weber}, were used by
Linde and Zelnikov\ \cite{LZ} to model a higher-dimensional universe in
which different numbers of dimensions inflate in different patches of the
universe. However, it was subsequently shown that this behaviour, like the
Weber solution, is unstable \cite{BY, sol}. We note that our new solutions
to gravity theories with $\beta \neq 0$ possess anisotropic inflationary
behaviour without requiring that the spatial curvature is positive and are
distinct from the Kantowski-Sachs phenomenon.

The Bianchi type solutions given above inflate in the presence of a positive
cosmological constant $\Lambda $. However, they are neither de Sitter, nor
asymptotically de Sitter; nor do they have initial singularities. Let us
examine how these models evade the conclusions of the cosmic no-hair
theorem. Specifically, consider the type $II$ solution, eq.(\ref{type2}).We
define the time-like vector $\mathbf{n}=\partial /\partial t$ orthogonal to
the Bianchi type $II$ group orbits, and introduce an orthonormal frame. We
define the expansion tensor $\theta _{\mu \nu }=n_{\mu ;\nu }$ and decompose
it into the expansion scalar, $\theta \equiv \theta _{~\mu }^{\mu }$ and the
shear, $\sigma _{\mu \nu }\equiv \theta _{\mu \nu }-(1/3)(g_{\mu \nu
}+n_{\mu }n_{\nu })$, in the standard way. The Hubble scalar is given by $%
H=\theta /3$. For the type $II$ metric, we find (in the orthonormal frame) 
\[
\theta =2b,\quad \sigma _{\mu \nu }=\frac{1}{6}\mathrm{diag}(0,2b,-b,-b).
\]%
As a measure of the anisotropy, we introduce dimensionless variables by
normalizing with the expansion scalar: 
\[
\Sigma _{\mu \nu }=\frac{3\sigma _{\mu \nu }}{\theta }=\mathrm{diag}\left( 0,%
\frac{1}{2},-\frac{1}{4},-\frac{1}{4}\right) .
\]%
Interestingly, the expansion-normalised shear components are constants (and
independent of the parameters $\alpha $, $\beta $, and $\Lambda $) and this
shows that these solutions violate the cosmological no-hair theorem (which
requires $\sigma _{\mu \nu }/\theta \rightarrow 0$ as $t\rightarrow \infty $%
). To understand how this solution avoids the no-hair theorem of, say, ref. 
\cite{wald}, rewrite eq.(\ref{field}) as follows: 
\[
G_{\mu \nu }=\widetilde{T}_{\mu \nu },\quad \widetilde{T}_{\mu \nu }\equiv
-\Lambda g_{\mu \nu }-\Phi _{\mu \nu }+\kappa T_{\mu \nu }.
\]%
In this form the higher-order curvature terms can be interpreted as matter
terms contributing a fictitious energy-momentum tensor $\widetilde{T}_{\mu
\nu }$. For the Bianchi $II$ solution we find 
\begin{eqnarray}
\widetilde{T}_{\mu \nu } &=&\tfrac{1}{4}\mathrm{diag}\left(
5b^{2}-a^{2},-3b^{2}+3a^{2},-7b^{2}-a^{2},-7b^{2}-a^{2}\right)   \nonumber \\
&=&\mathrm{diag}(\widetilde{\rho },\widetilde{p}_{1},\widetilde{p}_{2},%
\widetilde{p}_{3}).
\end{eqnarray}%
where $\widetilde{\rho }$ and $\widetilde{p}_{i}$ are the energy density and
the principal pressures, respectively. The no hair theorems require the
dominant energy condition (DEC) and the strong energy condition (SEC) to
hold. However, since $\tilde{\rho}+\tilde{p}_{1}+\tilde{p}_{2}+\tilde{p}%
_{3}=-3b^{2}<0$ the SEC is always violated when $b\neq 0$. The DEC is
violated when $\tilde{\rho}<0$ and the weak energy condition (WEC) is also
violated because $\tilde{\rho}+\tilde{p}_{2}=\tilde{\rho}+\tilde{p}%
_{3}=-(a^{2}+b^{2})/2<0.$These violations also ensure that the singularity
theorems will not hold for these universes and they have no initial or final
singularities.

Are these solutions stable? Due to the complexity of the equations of motion
it is difficult to extract information about the stability of these
non-perturbative solutions in general. In the class of spatially homogeneous
cosmologies the dynamical systems approach has been extremely powerful for
determining asymptotic states of Bianchi models. A similar approach can be
adopted to the class of models considered here; however, the complexity of
the phase space increases dramatically due to the higher-derivative terms.
Nonetheless, some stability results can be easily extracted. Consider, for
example, a perfect fluid with a barotropic equation of state, $p=w\rho $,
where $w$ is constant. Due to the exponential expansion, the value of the
deceleration parameter is $q\equiv -(1+\dot{H}/H^{2})=-1$ for the type $II$
and $VI_{h}$ solutions given. Hence, these vacuum solutions will be stable
against the introduction of a perfect fluid with $w>-1$. This includes the
important cases of dust ($w=0$), radiation ($w=1/3$) and inflationary
stresses ($-1<w<-1/3$).

For perturbations of the shear and the curvature, the situation is far more
complicated. Even within the class of Bianchi models in general relativity a
full stability analysis is lacking. However, in some cases, some of the
modes can be extracted. Consider again the Bianchi type $II$ solution, eq. (%
\ref{type2}). Using the trace of the evolution equations, eq.(\ref{tr}), we
consider a perturbation of the Ricci scalar: 
\[
R\approx 4\Lambda +r_{1}e^{\lambda _{1}t}+r_{2}e^{\lambda _{2}t}.
\]%
Using $\Box R=-(\ddot{R}+\theta \dot{R})$, which is valid for spatially
homogeneous universes, eq.(\ref{tr}) again implies, to lowest order: 
\[
\lambda _{1,2}=-\frac{3H}{2}\left( 1\pm \sqrt{1-\frac{2}{9H^{2}(3\alpha
+\beta )}}\right) ,
\]%
for $(3\alpha +\beta )\neq 0$. This shows that the perturbation of the Ricci
scalar gives the same eigenmodes for the anisotropic solutions of type $II$
and $VI_{h}$ as it did for perturbations of de Sitter spacetime in eq.(\ref%
{eig}). In order to determine the stability of other modes, like shear and
anisotropic curvature modes, further analysis is required.

\section{Discussion}

The solutions that we have found raise new questions about the thermodynamic
interpretation of spacetimes. We are accustomed to attaching an entropy to
the geometric structure created by the presence of a cosmological constant,
for example the event horizon of de Sitter spacetime. Do these
anisotropically inflating solutions have a thermodynamic interpretation? If
they are stable they may be related to dissipative structures that appear in
non-equilibrium thermodynamics and which have appeared been identified in
situations where de Sitter metrics appear in the presence of stresses which
violate the strong energy condition \cite{jb1, jb2, jb3, jb4}. They also
provide a new perspective on the physical interpretation of higher-order
gravity terms in the gravitational Lagrangian.

In summary: we have found exact cosmological solutions of a gravitational
theory that generalises Einstein's by the addition of quadratic curvature
terms to the action. These solutions display the new phenomenon of
anisotropic inflation when $\Lambda >0$. They do not approach the de Sitter
spacetime asymptotically and provide examples of new outcomes for inflation
that is driven by a $p=-\rho $ stress and begins from 'general' initial
conditions.

\section*{Acknowledgment}

SH was supported by a Killam Postdoctoral Fellowship.


\begin{thebibliography}{99}
\bibitem{inflate} A. Guth, Phys. Rev. D 23, 347 (1981); A.D. Linde, Phys.
Lett. B129, 177 (1983).

\bibitem{jdb} J.D. Barrow, Perturbations of a De Sitter Universe, In \textit{%
The Very Early Universe}, eds. G. Gibbons, S.W. Hawking and S.T.C Siklos,
(Cambridge UP, Cambridge, p.267, 1983).

\bibitem{bou} W. Boucher and G.W. Gibbons, In \textit{The Very Early Universe%
}, ed. G. Gibbons, S.W. Hawking and S.T.C Siklos, (Cambridge UP, Cambridge,
p. 273, 1983).

\bibitem{starob} A.A. Starobinskii, Sov. Phys. JETP Lett. 37, 66 (1983).

\bibitem{jss} L.G. Jensen and J. Stein-Schabes, Phys. Rev. D 35, 1146 (1987).

\bibitem{wald} R. Wald, Phys. Rev. D 28, 2118 (1983).

\bibitem{jb4} J.D. Barrow, Phys. Lett. B, 187, 12 (1987).

\bibitem{power} F. Lucchin and S. Mataresse, Phys. Rev. D 32, 1316 (1985);
L.F. Abbott and M.B. Wise, Nucl. Phys. B 244, 541 (1984).

\bibitem{inter} J.D. Barrow, Phys. Lett. B, 235, 40 (1990); J.D. Barrow and
P. Saich, Phys. Lett B 249, 406 (1990); J.D. Barrow and A.R. Liddle, Phys.
Rev. D 47, R5219 (1993); A.D. Rendall, Class. Quant. Grav. 22,1655 (2005).

\bibitem{barr} J.D. Barrow, Nucl. Phys. B 296, 697 (1988).

\bibitem{barrcot} J.D. Barrow and S. Cotsakis, Phys. Lett. B. 214, 515
(1988).

\bibitem{maeda} K.I. Maeda, Phys. Rev. D 39, 3159 (1989).

\bibitem{Berkin}
A. Berkin, Phys. Rev. D 44, 1020 (1991).

\bibitem{BCX}
E. Bruning, D. Coule, C. Xu, Gen. Rel. Grav. 26, 1197 (1994).

\bibitem{Schmidt}
H.-J. Schmidt, \texttt{gr-qc/0407095}.

\bibitem{eternal} A.D. Linde, Phys. Lett. B 175, 395 (1986); A. Vilenkin,
Phys. Rev. D 27, 2848 (1983).

\bibitem{BT} J.D. Barrow and F.J. Tipler, \textit{The Anthropic Cosmological
Principle}, (Oxford UP, Oxford, 1986).

\bibitem{DT} S. Deser and B. Tekin, \textit{Phys. Rev.} D67, 084009 (2003).

\bibitem{BO} J.D. Barrow and A.C. Ottewill, J. Phys. A 16, 2757 (1983).

\bibitem{st} V. M\"{u}ller, H.-J. Schmidt and A.A. Starobinskii, Phys. Lett.
B 202, 198 (1988).

\bibitem{bm} A. Berkin and K.I. Maeda, Phys. Rev. D 44, 1691 (1991).

\bibitem{gott} S. Gottl\"{o}ber, V. M\"{u}ller and A.A. Starobinskii, Phys.
Rev. D 43, 2510 (1991).

\bibitem{ss} A.A. Starobinskii and H.-J. Schmidt, Class. Quantum Grav. 4,
695 (1987).

\bibitem{gss} H.-J. Schmidt, Class. Quantum Grav. 5, 233 (1988).

\bibitem{bg} J.D. Barrow and G. G\"{o}tz, Phys. Lett. B 231, 228 (1989).

\bibitem{rend} A.D. Rendall, Annales Henri Poincar\'{e} A5,1041 (2004)

\bibitem{Faraoni}
V.  Faraoni, Phys. Rev. D72 061501 (2005).

\bibitem{FN}
V. Faraoni and S. Nadeau, Phys. Rev. D72 124005 (2005).

\bibitem{PSWZ} J. Patera, R.T. Sharp, P. Winternitz and H. Zassenhaus, J.
Math. Phys. 17, 986 (1976).

\bibitem{L1} J. Lauret, Math. Ann. 319, 715 (2001), Quart. J. Math. 52, 463
(2001), Math. Z. 241, 83 (2002), Diff. Geom. Appl. 18, 177 (2003).

\bibitem{SigTalk} S. Hervik, talk held at the CMS Summer 2004 meeting,
Halifax, NS, Canada.

\bibitem{jb1} J.D. Barrow, Phys. Lett. B 180, 335 (1986). \qquad

\bibitem{jb2} J.D. Barrow, Nucl. Phys. B 310, 743 (1988).

\bibitem{jb3} J.D. Barrow, Phys. Lett. B 183, 285 (1987).

\bibitem{Kaloper}
N. Kaloper, Phys. Rev. D 44, 2380 (1991).

\bibitem{weber} E. Weber, J. Math. Phys. 25, 3279 (1984).

\bibitem{LZ} A.D. Linde and M.I. Zelnikov, Phys. Lett. B 215, 59 (1988).

\bibitem{BY} J.D. Barrow and J. Yearsley, Class. Quantum Grav. 13, 2965
(1996).

\bibitem{sol} H.H. Soleng, Class. Quantum Grav. 6, 1387 (1989).
\end{thebibliography}
\end{document}